\documentclass[conference 10pt,twocolumn]{IEEEtran}

\newcommand{\qed}{\mbox{}\hspace*{\fill}\nolinebreak\mbox{$\rule{0.55em}{0.55em}$}} 

\newcommand{\T}{\textstyle}

\newcommand{\expect}{\mathbb{E}}
\newcommand{\prob}{\mathbb{P}}

\newtheorem{theorem}{Theorem}
\newtheorem{lemma}{Lemma}
\newtheorem{definition}{Definition}

\usepackage{amsmath, latexsym, amsfonts, amssymb}
\usepackage[mathscr]{eucal}
\usepackage{graphicx}

\begin{document}
\title{On the Asymptotic Behavior of Selfish Transmitters Sharing a Common Channel}

\author{\authorblockN{Hazer Inaltekin\authorrefmark{1}
, Mung Chiang \authorrefmark{1}, H. Vincent Poor\authorrefmark{1},
Stephen B. Wicker\authorrefmark{2},
}\\
\authorblockA{\authorrefmark{1}Department of Electrical Engineering,
Princeton University, Princeton, NJ 08544 \\
Email: \{hinaltek, mchiang, poor\}@princeton.edu} \\
\authorblockA{\authorrefmark{2} School of Electrical and Computer
Engineering, Cornell University, Ithaca,  14850 \\
Email: wicker@ece.cornell.edu} \thanks{This research was supported in part by the National Science Foundation under Grants ANI-03-38807 and CNS-06-25637.}}
\date{}
\maketitle
\begin{abstract}
\textbf{\small{This paper analyzes the asymptotic behavior of a
multiple-access network comprising a large number of \emph{selfish}
transmitters competing for access to a common wireless
communication channel, and having different utility
functions for determining their strategies.  A necessary and sufficient condition is given 
for the total number of packet 
arrivals from selfish transmitters to converge in distribution.  The asymptotic packet
arrival distribution at Nash equilibrium is shown to be a mixture of a Poisson distribution and finitely many Bernoulli distributions.}}
\end{abstract}

\section{Introduction}
\begin{figure}[t]
\centering{\includegraphics[scale=.35]{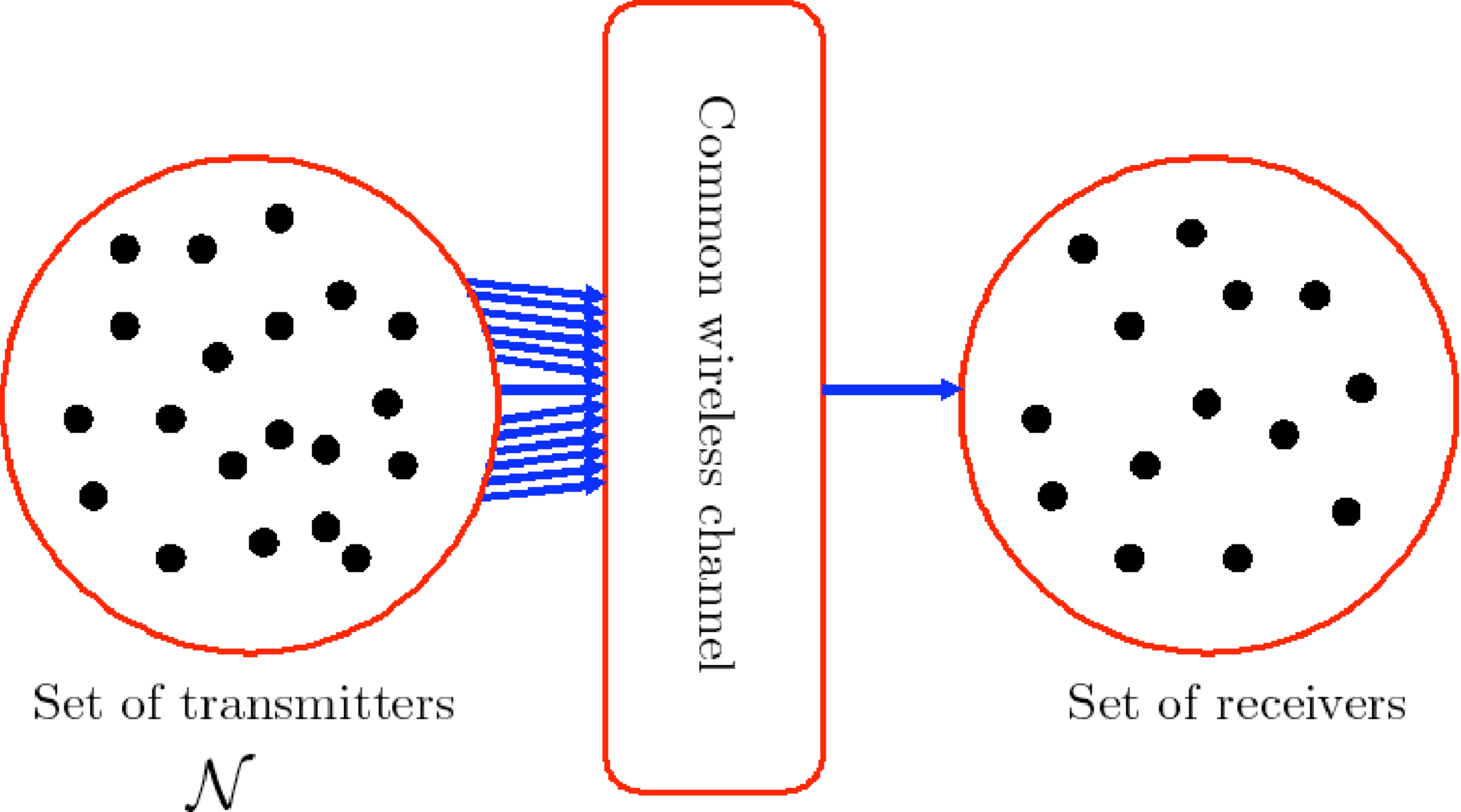}}\caption{Network
model in which $n$ selfish transmitters contend for the access of a
common wireless communication channel to communicate with their
intended receivers in the receiver set.} \label{Fig: network model}
\end{figure}

To investigate the behavior of a multiple-access communication
network consisting of large number of selfish transmitters, we
consider the network model depicted in Fig. \ref{Fig: network
model}.   In Fig. \ref{Fig: network model}, each transmitter in the
transmitter set  has an intended receiver in the receiver set.  In
the context of cellular networks, the transmitter set consists of mobile
users requesting uplink reservations to communicate with a base
station.  In a more general setting, it can be thought of as containing
some number of wireless transmitters that are closely located in a
wireless ad-hoc network, and that are willing to communicate with another
close-by node.  The results in this paper can be viewed as
characterizing the local behavior of dense wireless networks
containing selfish nodes and using a collision channel model at the medium access control
(MAC) layer.  The collision channel model has been extensively used in the
past (e.g., \cite{Collision Model: 1}, \cite{Collision Model:2}), and it is used to
characterize the behavior of networks using no power control and
containing nodes with single packet detection capabilities.  The
protocol model defined in \cite{Capacity: Gupta and Kumar} is a
variation of the collision model.

\subsection{Game Definition}
We assume that transmitter nodes always have packets to transmit, and a
transmission fails if there is more than one transmission at the same time.  The cost
of unsuccessful transmission of node $i$ is $c_i \in (0,\infty)$.  A likely meaning that can be attributed to $c_i$'s is the useless power expenditure caused by failed packets.  If
a transmission is successful, the node that transmitted its packet
successfully gets a normalized utility of $1$ unit.  We model this situation by using a
strategic game $G(n, \mathbf{c})$, which is defined formally as
follows: 
\emph{\begin{definition} \label{Def: One-shot game}
A heterogenous one-shot random access game with $n$ transmitter nodes
is the game $ G(n, \mathbf{c}) = \langle \mathcal{N},
(\mathcal{A}_{i})_{i \in \mathcal{N}}, (u_{i})_{i \in \mathcal{N}}
\rangle$ such that $\mathcal{N} = \{1,\ 2,\ldots,\ n\}$ is the set
of transmitters, $\mathcal{A}_i = \{0,1\}$ for all $i \in
\mathcal{N}$, where $\mathcal{A}_i$ is the set of actions for node $i$ and $1$ means transmission and $0$ means back-off,
$\mathbf{c} = (c_i)_{i \in \mathcal{N}}$ where $c_i$ is the cost of
unsuccessful transmission for node $i$, and the utility function
$u_i$ for all $i \in \mathcal{N}$ is defined as:
\begin{eqnarray} u_i(\mathbf{a}) &=& 0 \ \ \ \ \ \mbox{if} \ \ a_i = 0, \nonumber \\
u_i(\mathbf{a}) &=& 1 \ \ \ \ \
\mbox{if} \ \ \|\mathbf{a}\|_{l^1}=1 \ \ \mbox{and} \ \ a_i = 1, \nonumber \\
u_i(\mathbf{a}) &=&-c_i \ \ \mbox{if} \ \ \|\mathbf{a}\|_{l^1} \geq
2 \ \mbox{and} \ \ a_i = 1. \nonumber
\end{eqnarray}
\end{definition}}
In Definition \ref{Def: One-shot game}, $\|\cdot\|_{l^1}$ denotes the
$l^1$ norm for vectors in $\mathbb{R}^n$, and is thus the sum
of the absolute values of the components of a vector.  If $c_i = c>0$ for all $i \in
\mathcal{N}$, then we will denote $G(n,\mathbf{c})$ by $G(n,c)$,
and call it a \emph{homogenous} one-shot random access game.

\subsection{A Note on Notation}
$Po(\lambda)$ and $Bern(p)$ will indicate a Poisson distribution
with mean $\lambda$ and a 0-1 Bernoulli distribution with mean $p$,
respectively, as well as the generic random variables with these
distributions.  For any given two discrete distributions
$\mu$ and $\nu$ on the set of integers $\mathbb{Z}$, $d_V(\mu, \nu)$
denotes the \emph{variational distance} between them, which is defined as $d_V(\mu,\nu) = \sum_{z \in \mathbb{Z}} |\mu(z)-\nu(z)|$.  If $X$ and $Y$
are random variables with distributions $\mu$ and $\nu$, we
sometimes write $d_V(X,Y)$ in stead of $d_V(\mu,\nu)$ for ease
of understanding.  If one of the arguments of $d_V$ contains
a summation of some random variables, this refers to the convolution of
their respective distributions.  

If a sequence of probability
distributions $\{\mu_n\}_{n=1}^\infty$ converges (in the usual sense of convergence 
in distribution) to another probability distribution $\mu$, we represent this convergence by
$\mu_n \Rightarrow \mu$ as $n \rightarrow \infty$.  

As in standard terminology, we call a Nash
equilibrium a \emph{fully-mixed Nash equilibrium} (FMNE) if all of the
transmitters transmit with some positive probability in $(0,1)$ at this
equilibrium.  We call a Nash equilibrium a \emph{pure strategy Nash equilibrium} if all transmitters choose their actions deterministically.  Therefore, any given transmitter $i \in \mathcal{N}$ either transmits or backs-off with probability one at a pure strategy equilibrium.  

$X_i^{(n)}$ is the 0-1 random variable showing the
action chosen by transmitter $i \in \mathcal{N}$ when the game
$G(n,\mathbf{c})$ is played.  As noted before, $0$ means back-off, and $1$ means
transmit.  Let $p_{i,n}$ denote the transmission probability of
transmitter $i$ when there are $n$ transmitters contending for the
channel access.  Also let $S_n$ represent the total number of packet
arrivals when there are $n$ transmitters contending for the channel
access.  Note that $S_n = \sum_{i=1}^n X_i^{(n)}$.  For a given set
$\mathcal{N}_0$, $|\mathcal{N}_0|$ will represent the cardinality of
this set.

\subsection{Nash Equilibria of $G(n,\mathbf{c})$}
The transmission probability vector at which all nodes back-off with
probability one is not a Nash equilibrium of $G(n, \mathbf{c})$
since any node can obtain positive utility by setting its
transmission probability to a positive number given the fact that
others do not transmit.  Therefore, there is an incentive for nodes
to deviate from the strategy profile at which all of them back-off.
As a result, at Nash equilibria of $G(n, \mathbf{c})$, we expect to
observe some of the transmitters transmitting with some positive
probabilities and the remaining back-off with probability one.  To
further investigate this point, let $\pi: \bigcup_{n=2}^\infty
{\mathbb{R}}^n_+ \rightarrow \mathbb{R}_+$ be such that for any
$\mathbf{c} \in \bigcup_{n=2}^\infty \mathbb{R}^n_+$,
$\pi(\mathbf{c}) = \prod_i \frac{c_i}{1+c_i}$.  The following
theorem from \cite{Random Access Game: Inaltekin} characterizes the Nash
equilibria of this game.
\emph{\begin{theorem} 
\label{Thm: Nash Equilibria for inhomogenous M person game}
Let $X_i^{(n)}$ be the action chosen by transmitter $i \in
\mathcal{N}$, $\mathbf{c} \in \mathbb{R}^n$ and $\mathcal{N}_0
\subseteq \mathcal{N}$ with $2 \leq |\mathcal{N}_0| \leq n$. Then,
$G(n,\mathbf{c})$ has $n$ pure-strategy Nash equilibria.  Moreover,
any mixed-strategy profile such that nodes in $\mathcal{N}_0$
transmit with some positive probability, and nodes in
$\mathcal{N}-\mathcal{N}_0$ back-off with probability 1 is a Nash
equilibrium if and only if $ \T \prob\{X_i^{(n)} = 1\} = \T
1-(\frac{1+c_i}{c_i})(\pi(\mathbf{c}'))^\frac{1}{|\mathcal{N}_0|-1}$
for $i \in \mathcal{N}_0$, and $\frac{c_i}{1+c_i}
\stackrel{(\geq)}{>} \pi(\mathbf{c'})^\frac{1}{|\mathcal{N}_0|-1}$
for all $i \in \mathcal{N}$ (with $\geq$ if $i \in
\mathcal{N}-\mathcal{N}_0$), where $\mathbf{c}' = (c_i)_{i \in
\mathcal{N}_0}$.
\end{theorem}}

\subsection{Review: Homogenous Case}
We briefly mention the form of the asymptotic distribution of the
total number of packet arrivals when all transmitters have identical utility functions.  In this case, the
necessary and sufficient condition given in Theorem \ref{Thm: Nash
Equilibria for inhomogenous M person game} can be satisfied for any
subset $\mathcal{N}_0$ of $\mathcal{N}$ with $| \mathcal{N}_0| \geq 2$ for proper choice of the nodes' transmission probabilities.  Therefore, for any given $\mathcal{N}_0 \subseteq
\mathcal{N}$ with $|\mathcal{N}_0| \geq 2$, a mixed strategy Nash equilibrium at which
only the transmitters in $\mathcal{N}_0$ transmit with some positive
probability, and the rest of them back-off with probability one
exists.  At such a Nash equilibrium, the transmission probabilities of
transmitters in $\mathcal{N}_0$ are all equal to $p = 1 -
\big(\frac{c}{1+c}\big)^{\frac{1}{|\mathcal{N}_0|-1}}$.  Thus,
transmitters transmit with probability $p = 1 -
\big(\frac{c}{1+c}\big)^{\frac{1}{n-1}}$ at the FMNE.  Hence, at the
FMNE of the homogenous random access game, $S_n$ becomes a binomial
random variable with the success probability $p = 1 -
\big(\frac{c}{1+c}\big)^{\frac{1}{n-1}}$.  Since $n \cdot \big(1 -
\big(\frac{c}{1+c}\big)^{\frac{1}{n-1}}\big)$ approaches to
$-\log\big(\frac{c}{1+c}\big)$ as $n$ goes to infinity, $S_n$
converges, in distribution, to a Poisson distribution with mean
$-\log\big(\frac{c}{1+c}\big)$, which can be shown by using Poisson approximation
the binomial distribution (\cite{Feller: Probability 1}).  

Further details can be found in \cite{Random Access Game: Inaltekin}.  For
the rest of the paper, our aim is to prove the limit theorem
for $S_n$ in the more general case when nodes do not have identical utility functions.  
We first give a counter example showing that the
limiting distribution of $S_n$ cannot always be a pure Poisson
distribution.  In this latter case, we then, however, show that it can be arbitrarily
closely approximated in distribution by a summation of finitely many
independent Bernoulli random variables and a Poisson random
variable.

\subsection{Related Work}
Two closely related work are \cite{Allen: Multipacket} and \cite{Eitan: Aloha}.  In these work, they analyze the performance of Slotted ALOHA protocol with 
selfish transmitters by only considering the homogenous case where selfish nodes have identical utility functions.  Moreover, they do not provide any results regarding the 
asymptotic packet arrival distribution.  In \cite{Random Access Game: Inaltekin}, we mostly focused on the asymptotic channel throughput and the asymptotic packet arrival
distribution in the homogenous case for the same problem set-up.  We also provided a weaker necessary condition for the convergence of packet arrivals in distribution in the heterogeneous case. 
Different from the existing work in the literature, this paper will concentrate on the asymptotic packet arrival distribution in the more general case when selfish transmitters having different utility functions
contend for the access of a common wireless communication channel.  We provide a necessary and sufficient condition for the convergence of total number of packet arrivals in distribution as the number
of selfish transmitters increases to infinity.  We also specify the form of the asymptotic packet arrival distribution.    

\section{Limiting Behavior of $S_n$ in the Heterogenous Case}
We start our discussion with an example illustrating that the Poisson type convergence does not occur
in general in the heterogeneous case.  This result, while somewhat negative, will shed light on
the form of the limiting distributions for
$S_n$.  In this example, the limiting
distribution of the packet arrivals will be a mixture of a Poisson
distribution and several Bernoulli distributions.

\textbf{Example:} We consider the FMNE of the one-shot random access
game, and let $\mathbf{c}_n =
(M_1,M_2,\ldots,M_l,\underbrace{1,1,\ldots,1}_{n-l \
\mbox{\scriptsize of them}})$. By Theorem \ref{Thm: Nash Equilibria
for inhomogenous M person game}, $G(n,\mathbf{c}_n)$ has an FMNE if
and only if the following conditions are satisfied:
\begin{eqnarray} \lefteqn{\frac{M_i}{1+M_i} >
\left(\frac{1}{2}\right)^\frac{n-l}{n-1}\prod_{j=1}^l\left(\frac{M_j}{1+M_j}\right)^\frac{1}{n-1}
\ \mbox{for} \ 1\leq i\leq l,} \hspace{7.7cm} \label{eqn: example-II
N&S condition} \\ 
\lefteqn{\mbox{and}} \hspace{8cm} \nonumber \end{eqnarray}
\begin{eqnarray}
\lefteqn{(\frac{1}{2})^{l-1} > \prod_{j=1}^l \frac{M_j}{1+M_j} \
\mbox{for} \ l+1\leq i\leq n.} \hspace{7.7cm}
\end{eqnarray} 
Since the right-hand side of (\ref{eqn:
example-II N&S condition}) approaches to $\frac12$, we must choose $M_i > 1$ for
all $i \in \{1,2,...,l\}$ to have the FMNE for all $n$ large enough.
Any choice of $M_1,M_2,\ldots,M_l$ such that $M_i > 1$
for all $i \in \{1,2,...,l\}$ and $\prod_{j=1}^l\frac{M_j}{1+M_j} <
(\frac{1}{2})^{l-1}$ is good for our purposes.  One way of choosing
such $M_i$'s is to make all of the $\frac{M_i}{1+M_i}$'s smaller than
$(\frac{1}{2})^\frac{l-1}{l}$, which corresponds to $
M_1,M_2,\ldots,M_l \in
\left(1,\frac{1}{2^\frac{l-1}{l}-1}\right)$. 

For appropriately chosen $M_i$, $\ 1\leq i\leq l$, we have the
following transmission probabilities:
\begin{eqnarray} \T 
 p_{i,n} &=& \T 1 -
\frac{1+M_i}{M_i}\left(\frac{1}{2}\right)^\frac{n-l}{n-1}\prod_{j=1}^l\left(\frac{M_j}{1+M_j}\right)^\frac{1}{n-1}
\nonumber  \mbox{ for} \ 1\leq i\leq l, \nonumber \\ 
\mbox{and} \hspace{0cm} & & \nonumber \\
\T p_{i,n} &=& \T 1 -
2^\frac{l-1}{n-1}\prod_{j=1}^l\left(\frac{M_j}{1+M_j}\right)^\frac{1}{n-1}
\nonumber \mbox{ for} \ l+1\leq i\leq n. \nonumber
\end{eqnarray}
Define $Y_n = \sum_{i = l+1}^n X_i^{(n)}$. Then, $S_n = \sum_{i = 1}^l
X_i^{(n)} + Y_n$.  Observe that $p_{\max}^{(n)}
\stackrel{\bigtriangleup}{=} \max_{l+1\leq i\leq n}p_{i,n}
\rightarrow 0$ and 
\begin{eqnarray} \sum_{i=l+1}^n p_{i,n}
\rightarrow \log(2^{1-l}) + \sum_{j=1}^l
\log\left(1+\frac{1}{M_j}\right) \end{eqnarray} as $n \rightarrow \infty$.
Therefore, \begin{eqnarray} Y_n &\Rightarrow& Po\left(\log(2^{1-l}) +
\sum_{j=1}^l\log\left(1+\frac{1}{M_j}\right)\right), \hspace{1cm} \\
X_i^{(n)} &\Rightarrow& Bern\left(1-\frac{1+M_i}{2M_i}\right) \ \
\mbox{for} \ 1\leq i\leq l. \end{eqnarray}
As a result, we conclude, by using the continuity theorem and the
independence of the random variables $Y_n$ and $X_i^{(n)}$, that
\begin{eqnarray}S_n &\Rightarrow&
Po\left(\log(2^{1-l})+\sum_{i=1}^l\log\left(1+\frac{1}{M_i}\right)\right) \nonumber \\
& & \hspace{1.5cm} + \ \sum_{i=1}^l
Bern\left(1-\frac{1+M_i}{2M_i}\right). \label{eqn: Example II-Poisson
convergence}\end{eqnarray}

One interesting feature of this example is that we cannot find infinitely many $M_i$'s that
are uniformly bounded away from $1$, since
$\frac{1}{2^\frac{l-1}{l}-1} \rightarrow 1$ as $l \rightarrow
\infty$.  This observation will help us in obtaining the
asymptotic distribution of $S_n$ in the heterogeneous case.

For the rest of the paper, we focus on the asymptotic distribution
of $S_n$ at the FMNE of $G(n, \mathbf{c}_n)$ since the FMNE is the fairest Nash equilibrium at which all transmitters have a chance to transmit with some positive probability depending on their costs of failed transmissions.  More general results
can be found in \cite{Inaltekin: Thesis}.  Set $a_i =
\frac{c_i}{1+c_i}$ and $a_{\min}(n) = \min_{1\leq i \leq n}a_i$.  We
will assume that the costs of unsuccessful transmission of the nodes depend only on their internal parameters such as remaining battery
lifetime or energy spent per transmission.  Therefore, adding new
transmitters to the game does not change the costs of the
transmitters already playing the game.  Thus, $\alpha =
\inf_{i\geq 1} a_i = \lim_{n \rightarrow \infty} a_{\min}(n)$ is
well-defined.  The following two auxiliary results will help in
proving the main theorem, Theorem \ref{Thm: asymptotic arrivals in
inhomogneous case}, of the paper. Their proofs  can be found in \cite{Random Access Game: Inaltekin} and \cite{Inaltekin: Thesis}.
The first one states the convergence of the geometric mean of the
numbers $a_1, a_2, \ldots, a_n$ to a constant $\alpha
>0$ as $n \rightarrow \infty$ if the FMNE exists for all $n\geq 2$.
The second one states the convergence of the $a_i$'s to the same
constant $\alpha$ if the FMNE exists for all $n \geq 2$.

\begin{lemma} \label{Lemma: limit of geometric mean of a_i} \emph{Let $Geo(a_1,a_2,\ldots, a_n)$ denote the geometric mean of
$a_1, a_2, \ldots, a_n$.  If the FMNE exists for all $n \geq 2$
exists, then $\lim_{n \rightarrow \infty}Geo(a_1,\ldots,a_n)=\alpha
> 0$.}
\end{lemma}

\begin{lemma} \label{lemma: homogeneous case is the best possible case} \emph{If the FMNE exists for all $n \geq 2$,
then $\lim_{i \rightarrow \infty}a_i = \alpha$.}
\end{lemma}

\begin{figure}[t]
\centering{\includegraphics[scale=.4]{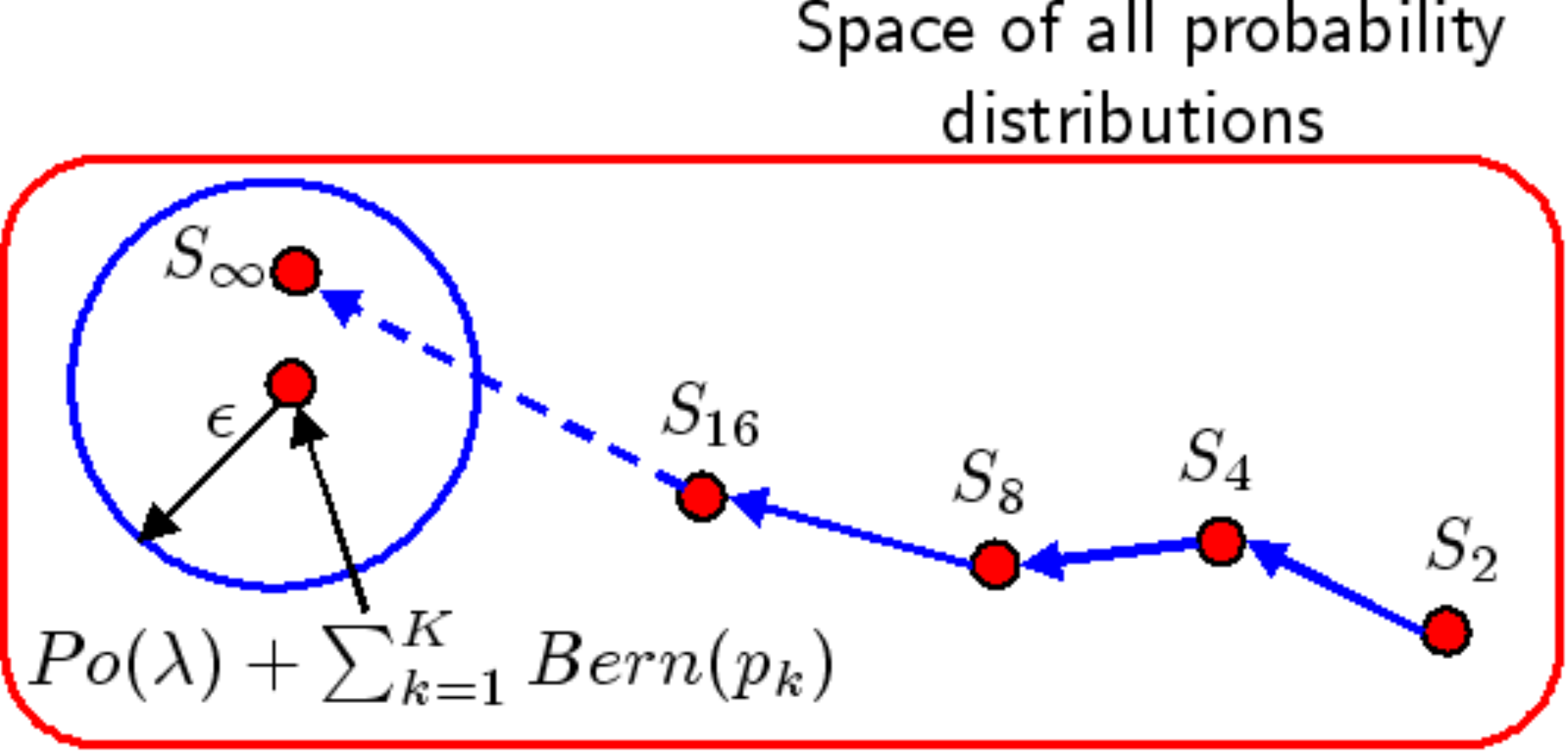}}\caption{A
pictorial explanation of Theorem \ref{Thm: asymptotic arrivals in
inhomogneous case}.  The limiting distribution of $S_n$ lies in a
small ball around the distribution of the random variable $
Po(\lambda)+\sum_{k=1}^K Bern(p_k)$.} \label{Fig: Theorem 2}
\end{figure}

In words, Lemma \ref{lemma: homogeneous case is the best possible
case} says that if the FMNE exists for all $n \geq 2$, then we can
find a $c>0$ such that for any given $\delta>0$, the costs of all the
transmitters, except for the finitely many of them, incurred as a
result of unsuccessful transmissions are concentrated in
$(c-\delta,c+\delta)$. Intuitively, we anticipate the selfish nodes
whose costs lie in $(c-\delta,c+\delta)$ to behave as in the
homogeneous case.  Thus, the total number of packet arrivals from these
nodes can be approximated by a Poisson random variable up to an
arbitrarily small error term $\epsilon(\delta)$ depending on
$\delta$.  The arrivals from the other finitely many nodes whose
costs lie outside of $(c-\delta,c+\delta)$ can be given by a summation of finitely many Bernoulli random variables.
Therefore, we expect that once $S_n$ converges in distribution, for
any given $\epsilon>0$, we should be able to find a Poisson random variable $Po(\lambda)$ and finitely many
Bernoulli random variables $\{Bern(p_j)\}_{j=1}^K$ such that $S_n$
can be approximated, in variational distance, by the sum of
$Po(\lambda)$ and $\{Bern(p_j)\}_{j=1}^K$ up to an error term less
than $\epsilon$.  A pictorial representation of this fact is
given in Fig. \ref{Fig: Theorem 2}.

The main result of the paper formally stating the above observation
is given in Theorem \ref{Thm: asymptotic arrivals in inhomogneous
case}.  In the proof of Theorem \ref{Thm: asymptotic arrivals in
inhomogneous case}, we let $p_{i,\infty} = \lim_{n\rightarrow
\infty} p_{i,n}$ when the FMNE exists for all $n \geq 2$.  Existence
of this limit can be shown by using Lemma \ref{Lemma: limit of
geometric mean of a_i}.

\begin{theorem} \label{Thm: asymptotic arrivals in inhomogneous case} \emph{ Assume FMNE exists for all
$n\geq 2$. Then, $S_n$ converges in distribution if and only if
$\lim_{n \rightarrow \infty} \sum_{i=1}^n p_{i,n} \in (0,\infty)$.
Moreover, for any $\epsilon>0$, there exists a Poisson random variable $Po(\lambda)$ and a collection of finitely many
Bernoulli random variables $\{Bern(p_k)\}_{k=1}^K$ such that
\begin{eqnarray}\limsup_{n \rightarrow \infty} d_V\left(S_n,
Po(\lambda)+\sum_{k=1}^K Bern(p_k)\right) \leq
\epsilon.\end{eqnarray}}
\end{theorem}

\emph{Proof:} $\Longleftarrow$: We first show the \textit{if}
direction. Suppose $\lim_{n \rightarrow \infty}
\sum_{i=1}^n p_{i,n} = m \in (0,\infty)$
exists. Let $m_n =\sum_{i=1}^n p_{i,n}$ and $S_n$ be distributed
according to $\mu_n$.  We will first show that
$\{\mu_n\}_{n=1}^\infty$ is a tight sequence of distributions. To
this end, we show that for each $\epsilon>0$, $\exists M \in \mathbb{N}$
such that $\prob\{S_n \in [0,M]\}\geq 1-\epsilon$.  Choose a
$\delta>0$ and choose $N \in \mathbb{N}$ large enough that $m_n \in
[m-\delta,m+\delta]$ for all $n \geq N$.  Then, by the Markov inequality,
\begin{eqnarray} \prob\{S_n >M\}
\leq \frac{\expect\left[(S_n-m_n)^2\right]}{(M-m_n)^2}. \nonumber
\end{eqnarray} 
We bound $\expect\left[(S_n-m_n)^2\right]$ as follows:
\begin{eqnarray} \expect\left[(S_n - m_n)^2\right] =
\sum_{i=1}^n Var\left(X_i^{(n)}\right)
\leq m_n \leq m+\delta. \nonumber \end{eqnarray} 
In addition,
$(M-m_n)^2 \geq (M-m-\delta)^2$.  Thus, \begin{eqnarray}\prob\{S_n >
M\} \leq \frac{m+\delta}{(M-m-\delta)^2}. \nonumber \end{eqnarray}
If $M$ is large enough, then we have $\prob\{S_n>M\}\leq \epsilon$ for
all $n \geq N$.  By making $M$ larger, if necessary, we have
$\prob\{S_n
> M\} = 0$ for all $n < N$.  As a result,
$\prob\left\{S_n \in [0,M]\right\} > 1- \epsilon \ \mbox{for all} \ n.$
Thus, $\{\mu_n\}_{n=1}^\infty$ is a tight sequence of distributions.
Now, we will show that $\mu_n$ converges, in variational distance, to a distribution
$\mu$.  This fact, combined with tightness of $\{\mu_n\}_{n=1}^\infty$, will imply that $\mu$ is in fact a probability distribution and $\mu_n
\Rightarrow \mu$.  By using Lemma \ref{Lemma: limit of geometric mean
of a_i} and Lemma \ref{lemma: homogeneous case is the best possible
case}, it can be shown that $\lim_{i \rightarrow \infty}p_{i,\infty}
= 0$.  Thus, for any given $\epsilon>0$, we can choose $K$ large
enough that $\max_{i \geq K}p_{i,\infty} \leq
\frac{\epsilon}{8m}$. 

Let $\lambda_n = \sum_{i=K}^n p_{i,n}$ and $\lambda = \lim_{n
\rightarrow \infty} \lambda_n$.  Then, by using the properties of variational distance, $d_V\left(S_n, Po(\lambda)+\sum_{i=1}^{K-1} Bern(p_{i,\infty})\right)$ 
can be bounded above as 
\begin{eqnarray} \lefteqn{d_V\left(S_n, Po(\lambda)+\sum_{i=1}^{K-1} Bern(p_{i,\infty})\right)} \hspace{8cm}
\nonumber \\  
\lefteqn{\leq
d_V\left(\sum_{i=1}^{K-1}X_i^{(n)},\sum_{i=1}^{K-1}
Bern(p_{i,\infty})\right)} \hspace{7.5cm} \nonumber \\
\lefteqn{+ \ d_V\left(Po(\lambda_n), Po(\lambda)\right)+ 2\sum_{i=K}^n
p_{i,n}^2} \hspace{7cm} \nonumber \\ 
\lefteqn{\leq
d_V\left(\sum_{i=1}^{K-1}X_i^{(n)},\sum_{i=1}^{K-1}
Bern(p_{i,\infty})\right)} \hspace{7.5cm} \nonumber \\
\lefteqn{+ \ d_V\left(Po(\lambda_n), Po(\lambda)\right)+ 2\max_{K\leq i
\leq n} p_{i,n} \sum_{i=K}^n p_{i,n}.} \hspace{7cm} \nonumber
\end{eqnarray}
Thus,
\begin{eqnarray} \lefteqn{\limsup_{n\rightarrow \infty}d_V\left(S_n,\sum_{i=1}^{K-1} Bern(p_{i,\infty}) +
Po(\lambda)\right)} \hspace{8cm} \nonumber \\
\lefteqn{ \leq
2\limsup_{n\rightarrow\infty}\big(\lambda_n.\max_{K\leq i \leq
n} p_{i,n}\big)} \hspace{7cm} \nonumber \\
\lefteqn{= 2\lambda \limsup_{n \rightarrow \infty} \max_{K\leq i
\leq n}p_{i,n} \ \ (\mbox{since} \ \lambda_n \rightarrow \lambda).}
\hspace{7cm} \nonumber
\end{eqnarray}
Let $i(n)$ be such that $p_{i(n),n} = \max_{K\leq i\leq n}p_{i,n}$.
Then, there exists a subsequence $\{n_k\}_{k=1}^\infty$ such that
\begin{eqnarray}\lim_{k \rightarrow \infty} p_{i(n_k),n_k} =
\limsup_{n \rightarrow \infty} \max_{K\leq i\leq n}p_{i,n}.\nonumber
\end{eqnarray} If $\{i(n_k)\}_{k=1}^\infty$ is a bounded sequence,
there exists a further subsequence $\{i(n_{k_j}\}_{j=1}^\infty$ such
that $\lim_{j \rightarrow \infty} i(n_{k_j}) = i^{**}$.  Since we are
considering a sequence of integers converging to another integer,
there exists $N \in \mathbb{N}$ such that we have $i(n_{k_j}) =
i^{**}$ for all $j\geq N$.  Thus,
\begin{eqnarray} \limsup_{n \rightarrow \infty} \max_{K\leq i\leq
n}p_{i,n} &=& p_{i^{**},\infty} \leq \max_{i\geq K}p_{i,\infty}.
\nonumber
\end{eqnarray}
If $\{i(n_k)\}_{k=1}^\infty$ is not a bounded sequence, then there exists
a further subsequence $\{i(n_{k_j})\}_{j=1}^\infty$ such that
$\lim_{j \rightarrow \infty} i(n_{k_j}) = \infty$.  Let $\gamma_n =
(\prod_{i=1}^n a_i)^\frac{1}{n-1}$.  Observe that transmission
probabilities at FMNE can be given as $p_{i,n} =
1-a_i^{-1}\gamma_n$.  So,
\begin{eqnarray} \limsup_{n \rightarrow \infty} \max_{K\leq i \leq
n} p_{i,n} &=& \lim_{j \rightarrow \infty} p_{i(n_{k_j}),n_{k_j}}
\nonumber \\ &=& 1 - \lim_{j \rightarrow
\infty}a_{i(n_{k_j})}^{-1}\lim_{j \rightarrow \infty}
\gamma_{n_{k_j}} \nonumber \\ &=& 0 \leq \max_{i \geq K}
p_{i,\infty}. \nonumber
\end{eqnarray} 
Therefore,
\begin{eqnarray} \lefteqn{\limsup_{n \rightarrow \infty}
d_V\left(S_n,Po(\lambda)+\sum_{i=1}^{K-1} Bern(p_{i,\infty})\right)}
\hspace{8cm} \nonumber \\ \lefteqn{\leq 2\lambda \max_{i \geq
K}p_{i,\infty} \leq \frac{\epsilon}{4}.} \hspace{6cm} \nonumber
\end{eqnarray} 
Thus, $\exists N \in \mathbb{N}$ large enough so that
\begin{eqnarray}d_V\left(S_n,Po(\lambda) + \sum_{i=1}^{K-1} Bern(p_{i,\infty})\right) \leq
\frac{\epsilon}{2} \nonumber \end{eqnarray} for all $n \geq N$.  As
a result, we conclude that $\{\mu_n\}_{n=1}^\infty$ is a Cauchy
sequence with respect to the metric $d_V$ on the set of all probability
measures $\mathcal{Z}$ on $\mathbb{Z}$.  This also implies that
$\{\mu_n(z)\}_{n=1}^\infty$ is a Cauchy sequence for all $z \in
\mathbb{Z}$, and therefore, converges for any $z \in \mathbb{Z}$.
Let $\mu(z) = \lim_{n\rightarrow \infty} \mu_n(z)$ for all $z \in
\mathbb{Z}$.  This combined with the tightness of $\{\mu_n\}_{n=1}^\infty$ implies
that $\mu$ is a probability measure and $\mu_n \Rightarrow \mu$.

$\Longrightarrow$: Now, we prove the \textit{only if} part. In fact,
this will be a general result for any sequence of triangular arrays
of Bernoulli random variables.  Suppose now that there exists an
$\mathbb{R}$ valued random variable $S_\infty$ such that $S_n$
converges in distribution to $S_\infty$.  First, assume
\begin{eqnarray}\limsup_{n \rightarrow \infty}m_n = \infty, \nonumber
\end{eqnarray} and let $Y_i^{(n)} = X_i^{(n)} - p_{i,n}$.  Set $R_n =
\sum_{i=1}^n Y_i^{(n)}$.  Consider $\expect\left[e^{-tY_i^{(n)}}\right]$ for
$t>0$.  We have
\begin{eqnarray}
\lefteqn{\expect\left[e^{-tY_i^{(n)}}\right]} \hspace{8cm} \nonumber \\
\lefteqn{\leq 1+
\frac{1}{2!}t^2\left|\expect\left[(Y_i^{(n)})^2\right]\right| +
\frac{1}{3!}t^3\left|\expect\left[(Y_i^{(n)})^3\right]\right| + \cdots.} \hspace{7.5cm}
\nonumber
\end{eqnarray} 
We will show $\left|\expect[(Y_i^{(n)})^k]\right| \leq p_{i,n}$ for
all $k$.  For $k=2$, 
\begin{eqnarray} \expect\left[(Y_i^{(n)})^2\right] =
Var\left(X_i^{(n)}\right) 
\leq \expect\left[\left(X_i^{(n)}\right)^2\right] = p_{i,n}. \nonumber \end{eqnarray}
For any $k\geq 3$, we have \begin{eqnarray}
\lefteqn{\left|\expect\left[\left(Y_i^{(n)}\right)^k\right]\right|}
\hspace{8cm} \nonumber \\
\lefteqn{\leq
\left(\expect\left[\left|Y_i^{(n)}\right|^{2k-2}\right]\right)^\frac{1}{2}\left(\expect\left[\left|Y_i^{(n)}\right|^2\right]\right)^\frac{1}{2}
\ \mbox{(H\"{o}lder's Ineq.)}} \hspace{7.5cm} \nonumber \\
\lefteqn{\leq
\left(\expect\left[\left|Y_i^{(n)}\right|^2\right]\right)^\frac{1}{2}
\left(\expect\left[\left|Y_i^{(n)}\right|^2\right]\right)^\frac{1}{2} \leq
p_{i,n}.} \hspace{7.5cm} \nonumber\end{eqnarray} 
Thus,
\begin{eqnarray}\expect\left[e^{-tY_i^{(n)}}\right] \leq 1+\frac{1}{2!}t^2p_{i,n} +
p_{i,n}\left(\frac{t^3}{3!} + \frac{t^4}{4!} + \cdots \right).\end{eqnarray}
Then, there is a $\delta_1>0$ such that, uniformly over all $p_{i,n}$, we
have
\begin{eqnarray} \expect\left[e^{-tY_i^{(n)}}\right] &\leq& 1+t^2p_{i,n} \ \
\mbox{for all}\ t \in (0,\delta_1). \nonumber \end{eqnarray} 
Now, make $\delta_1$ smaller (if necessary) so that $t^2 \leq \frac{t}{4}$.
Then, for $t \in (0,\delta_1)$, we have
\begin{eqnarray} \lefteqn{\prob\left\{S_n \leq \frac{m_n}{2}\right\}
\leq \frac{\expect\left[e^{-tR_n}\right]}{e^{\frac{tm_n}{2}}}
\ \ \mbox{(Markov Inequality)}} \hspace{8cm} \nonumber \\
\lefteqn{= e^{\frac{-tm_n}{2}}\prod_{i=1}^n\expect\left[e^{-tY_i^{(n)}}\right]}
\hspace{7.5cm} \nonumber \\  \lefteqn{\leq
e^{\frac{-tm_n}{2}}\prod_{i=1}^n(1+t^2p_{i,n})} \hspace{7.5cm}
\nonumber \\  
\lefteqn{=
\exp\left(\sum_{i=1}^n\left(\frac{-tp_{i,n}}{2}+\log\left(1+t^2p_{i,n}\right)\right)\right)}
\hspace{7.5cm} \nonumber \\
\lefteqn{\leq
\exp\left(\sum_{i=1}^n\left(\frac{-tp_{i,n}}{2}+t^2p_{i,n}\right)\right) \
\mbox{(since $\log(x)\leq x-1$)}} \hspace{7.5cm} \nonumber \\
 \lefteqn{\leq
\exp\left(\frac{-t}{4}\sum_{i=1}^{n}p_{i,n}\right).} \hspace{7.5cm}
\nonumber
\end{eqnarray}
Since $\limsup_{n\rightarrow \infty}m_n = \infty$, we can find a
subsequence of $\{m_n\}_{n=1}^\infty$, which we call
$\{m_{n(k)}\}_{k=1}^\infty$, such that $m_{n(k)}\geq k$.  Then,
$\prob\big\{S_{n(k)}\leq \frac{m_{n(k)}}{2}\big\}\leq
\exp\big(\frac{-tk}{4}\big)$.   On setting $\mathcal{A}_k = \{\omega \in
\Omega: S_{n(k)}(\omega) \leq \frac{m_{n(k)}}{2}\}$, we have
\begin{eqnarray} \sum_{k=1}^\infty \prob\left(\mathcal{A}_k\right)< \infty. \nonumber \end{eqnarray} 
By the Borel-Cantelli lemma,
$\prob\{\mathcal{A}_k \ i.o\}=0$. Thus, $\lim_{k\rightarrow
\infty}S_{n(k)} = \infty$ w.p.1.  Since almost sure convergence
implies convergence in distribution, we have $S_\infty = \infty$
w.p.1, which is a contradiction.  Thus, $\limsup_{n\rightarrow
\infty} m_n< \infty$.  In this case, we can find $C<\infty$ such that
$m_n\leq C$ for all $n$.  Now, we will show that
$\{S_n\}_{n=1}^\infty$ is uniformly integrable.
\begin{eqnarray} \expect\left[S_n^2\right] &=& \expect\left[\sum_{i=1}^n \left(X_i^{(n)}\right)^2\right] +
\expect\left[\sum_{i,j = 1 \atop i\neq j}^n X_i^{(n)}X_j^{(n)}\right]
\nonumber \\
&\leq& \sum_{i=1}^np_{i,n} + \sum_{i,j=1}^np_{i,n}p_{j,n} \nonumber
\\ &=& m_n + m_n^2 \leq C\left(1+C\right) < \infty. \nonumber \end{eqnarray}
Therefore, $\{S_n\}_{n=1}^\infty$ is uniformly integrable.  By
Skorohod's representation theorem, there exists a probability space
$\left(\Omega^\prime, \mathcal{F}^\prime, P^\prime\right)$ and random variables
$S_n^\prime$ and $S_\infty^\prime$ having the same distributions as 
$S_n$ and $S_\infty$, respectively, such that $S_n^\prime
\rightarrow S_\infty^\prime$ w.p.1.  By uniform integrability, we
also have $L^1$ convergence, i.e.,
\begin{eqnarray} \lim_{n\rightarrow \infty}\expect\left[S_n^\prime\right] =
\expect\left[S_\infty^\prime\right].\end{eqnarray} 
Therefore,
$\lim_{n\rightarrow \infty}m_n$ exists and belongs to $(0,\infty)$.
\qed

\section{Conclusion}
In this paper, we have analyzed the asymptotic behavior of
multiple-access networks containing large numbers of selfish
transmitters that share a common wireless communication channel to
communicate with their intended receivers.  In particular, we have
focused on the asymptotic distribution of the total number of packet
arrivals to the common wireless channel coming from these selfish
transmitters.  When selfish transmitters are identical to one
another in their utility functions, we have shown that the asymptotic
distribution of the total number of packet arrivals becomes equal to
a Poisson distribution.  On the other hand, when selfish
transmitters do not have identical utility functions, we have first
obtained a necessary and sufficient condition for the total number
of packet arrivals to converge in distribution. We then have shown that
the asymptotic packet arrival distribution can be arbitrarily
closely approximated in distribution by a summation of finitely many independent Bernoulli random variables and an independent Poisson random
variable.

{}

\end{document}